\documentclass[]{article}

\usepackage[margin=0.75in]{geometry}
\usepackage[capposition=top]{floatrow}
\usepackage{xcolor}
\usepackage{amsmath}
\usepackage{amsfonts}
\usepackage{graphicx}
\usepackage{float}
\usepackage{hyperref}
\usepackage{setspace}
\usepackage{multicol}
\usepackage{lipsum}
\usepackage{algorithm}
\usepackage{algpseudocode}
\usepackage{pythonhighlight}

\newcommand{\edit}[1]{\color{black}#1\color{black}}

\title{Estimating equations for causal survival analysis with pooled logistic regression}
\author{Paul N Zivich\textsuperscript{a,*}, Stephen R Cole\textsuperscript{a}, Bonnie E Shook-Sa\textsuperscript{b,c}, Justin B DeMonte\textsuperscript{b}, Jessie K Edwards\textsuperscript{a}}
\date{%
	\textsuperscript{a}Department of Epidemiology, Gillings School of Global Public Health, University of North Carolina at Chapel Hill, Chapel Hill, NC, USA\\%
	\textsuperscript{b}Department of Biostatistics, Gillings School of Global Public Health, University of North Carolina at Chapel Hill, Chapel Hill, NC, USA\\%
	\textsuperscript{c}Nuffield Department of Population Health, University of Oxford, Oxford, UK \\
	\textsuperscript{*}Corresponding Author. 135 Dauer Drive. Chapel Hill, NC, USA 27599. pzivich@unc.edu  \\%
	~\\
	\today
}

\begin{document}

\maketitle

\begin{abstract}
	\noindent
	\textbf{Background}: Pooled logistic regression models are commonly applied in survival analysis. However, the standard implementation can be computationally demanding, which is further exacerbated when using the nonparametric bootstrap for inference. To ease these computational burdens, investigators often coarsen time intervals or assume a parametric \edit{functional form } for time. These approaches impose restrictive assumptions, which may not always have a well-motivated substantive justification. \\
	\textbf{Methods}: Here, the pooled logistic regression model is re-framed using estimating equations to simplify computations and allow for inference via the empirical sandwich variance estimator, thus avoiding the more computationally demanding bootstrap. The proposed implementation is demonstrated using two examples with publicly available data. The performance of the empirical sandwich variance estimator is illustrated using a Monte Carlo simulation study. \\
	\textbf{Results}: As shown in the applied examples, the proposed implementation substantially reduced run-times and could be applied without needing to coarsen the data. In the simulation study, the empirical sandwich variance estimator results in nominal confidence interval coverage. \\
	\textbf{Conclusions}: The implementation proposed here offers a \edit{less computationally demanding } alternative to the standard implementation of pooled logistic regression without needing to impose restrictive constraints on time.
\end{abstract}


\newpage

\section{Introduction}

Causal survival analysis, as with analyses of time-to-event data more generally, must contend with right censoring. When the target parameter is the intent-to-treat effect (i.e., comparison of the cumulative density, or risk, if everyone had initiated treatment at baseline versus no one had initiated treatment at baseline), g-computation and inverse probability weighting (IPW) estimators have been proposed that address both confounding and informative right censoring in the observational setting \cite{hernanObservationPlansLongitudinal2008, hernanHazardsHazardRatios2010, howeSelectionBiasDue2016}. Each of these approaches relies on modeling some process over time (i.e., outcome process for g-computation, censoring process for IPW). Statistical models for these processes are commonly referred to as \textit{nuisance} models, since the parameters of these models are not of scientific interest themselves but are needed to estimate the target parameter(s). While a variety of different techniques can be used for the nuisance models \edit{in causal survival analysis } \cite{makuchAdjustedSurvivalCurve1982, chattonGcomputationDoublyRobust2022, edwardsSemiparametricGcomputationSurvival2024}, using pooled logistic regression with discretized time intervals is a popular choice \cite{hernanHazardsHazardRatios2010, hernanMarginalStructuralModels2001, matthewsComparingEffectEstimates2021, murrayAdherenceadjustmentPlacebocontrolledRandomized2020, murrayCausalSurvivalAnalysis2021a, smithEmulationTargetTrial2022, wanisEffectOpioidEpidemic2019, wanisAdjustingAdherenceRandomized2020, wenParametricGformulaImplementations2021, demonteAssessingCOVID19Vaccine2024}.

The standard implementation of a pooled logistic regression model uses a data set where a row corresponds to an interval of time (e.g., day, week, month) for an individual \cite{abbottLogisticRegressionSurvival1985, cupplesComparisonBaselineRepeated1988, dagostinoRelationPooledLogistic1990}. This data set is commonly referred to as a `long' data set since it includes multiple rows for the same person, one for each time period. These person-period observations are then used to fit a single logistic regression model which conditions on survival up to a given time point. As a rule of thumb, when the risk of the modeled event (e.g., outcome, censoring) is less than 10\% in any interval, the exponentiated estimates from this model can be interpreted as approximations of hazard ratios from a Cox proportional hazards model \cite{murrayCausalSurvivalAnalysis2021a}. While sometimes claimed otherwise \cite{murrayCausalSurvivalAnalysis2021a}, the \edit{model-based variance } estimator for the coefficients of this model is valid for inference with non-repeating (i.e., absorbing) events. While it first seems that because there are multiple rows per person that one needs to correct for this dependence \edit{with a cluster-robust variance estimator}, it can be shown that the likelihood factors into independent person-period probabilities and thus the \edit{model-based } variance estimator that assumes independent observations remains valid \cite[pg.~246-247]{allisonSurvivalAnalysisUsing2010}. However, variance estimation is more complicated when using pooled logistic regression as a nuisance model, as done with g-computation. \edit{When estimating the risk under a given treatment plan using g-computation, } the covariate-conditional risk is predicted from the estimated pooled logistic model and then marginalized over the observed baseline covariate distribution \cite{hernanHazardsHazardRatios2010}. Inference for the marginal risk thus depends on the sampling variability in the pooled logistic regression model and the \edit{model-based } variance estimates from the nuisance model can no longer be directly used for inference. To account for the uncertainty in the nuisance model with g-computation, the nonparametric bootstrap is a common approach \cite{hernanHazardsHazardRatios2010, edwardsSemiparametricGcomputationSurvival2024}. 

Use of pooled logistic regression for nuisance models with the standard approach faces several challenges. First, the size of the long data set can quickly become unwieldy, particularly when there are a large number of observations or a large number of time intervals. To reduce the number of records in the long data set, it is common practice to coarsen the time intervals (e.g., weeks instead of days) but coarsening leads to a loss of information \cite{edwardsSemiparametricGcomputationSurvival2024}. Second, variance estimation with the bootstrap is computationally demanding and can lead to substantial run-times even with modern computing infrastructures, as the pooled logistic regression models must be re-fit hundreds or thousands of times. Third, errors can easily be introduced when transforming the data set into a long data set (e.g., miscoding events, incorrectly aligning time). While these can be avoided through careful coding practices, they represent a potential failure point. These computational challenges have downstream impacts on scientific applications. As noted in the open-source code accompanying Wanis et al. \cite{wanisAdjustingAdherenceRandomized2020}, inference based on the bootstrap can take several hours to run. Alternatively, computational complexity may restrict the type of analyses considered, as noted in the supplementary materials of McConeghy et al. \cite{mcconeghyInfectionsHospitalizationsDeaths2022} Therefore, \edit{these computation issues pose a threat to the adoption and reproducibility of methods like g-computation. } Improvements to the computation of pooled logistic regression offer a way to materially improve causal survival analysis.

To \edit{enhance } the standard implementation of pooled logistic regression, we propose an alternative based on estimating equations. \edit{The proposed implementation makes two improvements. First, } the estimating equations avoid the need to create a long data set. \edit{Second, the target parameter and nuisance parameters can be jointly estimated so that } the variance of the target parameter can be consistently estimated \edit{using } the empirical sandwich variance estimator. This approach to variance estimation means that the pooled logistic model only needs to be fit once \edit{with the g-computation estimator}, unlike the bootstrap. 

The remainder of this article is organized as follows. In Section 2, a setup for the general estimation problem is provided. Section 3 reviews pooled logistic regression models and Section 4 expresses these models as estimating equations. Two examples using publicly available data are described in Section 5 along with corresponding Python code. Section 6 explores performance of the estimating equation approach with a Monte Carlo simulation. Finally, Section 7 summarizes the key messages and points towards areas of future work.

\section{Setup}

Let $T_i^a$ denote the potential event time under the baseline action $a \in \{0,1\}$ and $A_i$ denote the observed action for unit $i$. Here, the target interest is the intent-to-treat causal risk difference, or average causal effect, of $A$ on the event at time $t$, expressed as $F^1(t) - F^0(t) = E[I(T_i^1 \le t)] - E[I(T_i^0 \le t)]$, where $I(\cdot)$ is the indicator function and $t \in (0, \tau]$ where $\tau$ indicates the maximum follow-up time. The first task is to illustrate how the target parameter can be expressed in terms of the observed data, referred to as identification.

To start, suppose there is no censoring. Therefore, the observed data contains $n$ iid copies of $(W_i, A_i, T_i)$, where $T_i$ is the event time and $W_i$ is a vector of baseline covariates. Assume causal consistency (i.e., if $A_i = a$ then $T_i = T_i^a$), conditional exchangeability (i.e., $T^a \amalg A \mid W$), and positivity (i.e., $\Pr(A=a \mid W=w) > 0$ for $a \in \{0,1\}$ and $f_W(w) > 0$, where $f_W$ is the probability density function of $W$). Then it follows that 
\begin{equation}
	F^a(t) = E[E(I(T \le t) \mid A=a, W)] \text{ for } a \in \{0,1\}
	\label{id_result}
\end{equation}
\textit{Proof}:
\begin{equation*}
		E[E(I(T \le t) \mid A=a, W)] = E[E(I(T^a \le t) \mid A=a, W)] = E[E(I(T^a \le t) \mid W)] = E[I(T^a \le t)]
\end{equation*}
where the first equality follows from causal consistency, the second follows from conditional exchangeability with positivity by $W$, and the third follows from the law of iterated expectations. 

With censoring, $T_i$ is not observed for all units. Instead the observed data consists of $O_i = (W_i, A_i, T_i^*, \Delta_i)$, where $T_i^* = \min(T_i, C_i)$, $C_i$ is the censoring time, and $\Delta_i = I(T_i = T_i^*)$. Under the assumption that the hazards of the event and censoring are independent conditional on $A,W$ (with a corresponding positivity assumption), $E(I(T \le t) \mid A=a, W)$ in Equation \ref{id_result} can be expressed in terms of $T^*$ and $\Delta$ \cite{robinsCorrectingNoncomplianceDependent2000a}. 

These previous identification results immediately suggest a g-computation type estimator for the target parameter. Namely, a model is assumed for the baseline covariate-conditional probability of the outcome by time $t$, denoted as $\mu_a(t, W_i; \beta)$. This model is then used to predict the conditional probability of the outcome for each unit under each set value of $A$. These unit-specific predictions are then averaged. This g-computation estimator can be written as
\begin{equation}
	\hat{F}^a(t) = n^{-1} \sum_{i=1}^{n} \mu_a(t, W_i; \hat{\beta})
	\label{gcomp_estr}
\end{equation}
where $\mu_a(t, W_i; \hat{\beta})$ is estimated from a model for $E(T \le t \mid A, W)$. Therefore, the estimator for the \edit{causal risk difference } is $\hat{F}^1(t) - \hat{F}^0(t)$. A variety of different models can be used for $\mu_a(t, W_i; \beta)$ \cite{edwardsSemiparametricGcomputationSurvival2024}. Hereafter, we demonstrate estimation of \ref{gcomp_estr} using pooled logistic regression. 

For statistical inference under a population model, the variance of the target parameter estimator, $V\left(\hat{F}^1(t) - \hat{F}^0(t)\right)$, must be consistently estimated. Here, the variance of the \edit{causal risk difference } estimator depends on the uncertainty in the estimated nuisance model parameters, $\hat{\beta}$. Prior work has relied on a nonparametric bootstrap procedure \cite{murrayCausalSurvivalAnalysis2021a}. Briefly, one resamples the data with replacement and estimates $\beta$ and $F^1(t) - F^0(t)$ using the resampled observations. Rather than resampling, one could also consider the fractional-random-weight bootstrap. Regardless, either process is repeated a large number of times (e.g., 500). The collection of different estimates of $F^1(t) - F^0(t)$ can then be used to compute Wald-type 95\% Confidence Intervals (CIs) by taking their standard deviation as an estimate of the standard error of the target parameter estimator. Alternatively, one could also use the 2.5\textsuperscript{th} and 97.5\textsuperscript{th} percentiles for the confidence intervals, but this approach generally requires more iterations.

In addition to the \edit{causal risk difference } at time $t$, one might also be interested in how the \edit{risk difference } changes over time, i.e., $\{F^1(s) - F^0(s) : s \in (0, \tau]\}$. Implicitly, this is the case when risks or the risk difference is plotted across time. While the previous identification proof is sufficient to identify this function and estimation can also be done using the described g-computation estimator, inference is \edit{more } complicated. For functions (or parameter vectors), point-wise Wald-type CIs are no longer appropriate for inference, as valid point-wise CIs only cover at their nominal level for the point they correspond to. Therefore, their use when plotting effects across time can understate the uncertainty regarding how the function changes over $(0, \tau]$. Instead, Confidence Bands (CBs) for statistical inference are preferred, as they provide the advertised coverage for a function \cite{zivichConfidenceRegionsMultiple2026}.

\section{Pooled Logistic Regression}

Here, we review how the standard implementation of pooled logistic regression is applied. The first step in a standard implementation of pooled logistic regression involves coarsening time into discrete intervals. Let the duration of follow-up, $\mathcal{T}^* = (0, \tau]$, be divided into $K$ equally-spaced time intervals. Each interval $k \in \mathcal{K} = \{1, ..., K\}$ is defined as $(s_{k-1}, s_k]$, where $s_{k-1}$ and $s_k$ are the lower and upper limits for interval $k$ with $s_0 = 0$ and $s_K = \tau$.  While coarsening can result in a loss of granularity, choosing the intervals at the same resolution as the measurement of time in a given study results in no such losses. These intervals are used to construct a long data set, in which each row corresponds to a single participant for a single time interval (i.e., person-period observation) up till their final time under observation. Therefore, a person has a row for each time interval they contribute to. An event indicator is constructed which indicates whether a unit has the event in the interval $(s_{k-1}, s_k]$, i.e., $Y_{i,k} = I(s_{k-1} < T_i^* \le s_k) \Delta_i$. 

A logistic regression model for the probability of $Y_{i,k}$ is then fit to this long data set, conditional on survival to the start of interval $k$, any covariates ($A,W$ here), and some function of time. This pooled logistic regression model can be expressed as 
\begin{equation}
	\Pr(Y_{i,k} = 1 \mid A_i, W_i, T_i^* > s_{k-1}; \beta) = \text{expit}(\mathbb{X}_i \beta_X + \mathbb{S}_{i,k} \beta_S)
	\label{PLRM}
\end{equation}
where \edit{$\beta = (\beta_X^T, \beta_S^T)^T$} are unknown regression parameters, $\mathbb{X}_i$ is \edit{a row vector corresponding to } row $i$ of the design matrix for $g(A_i,W_i)$ with $g$ denoting a general row vector function, $\mathbb{S}_{i,k}$ is \edit{a row vector corresponding to } row $i$ of the design matrix for $s_k$, and $\text{expit}(a) = (1 + \exp(-a))^{-1}$. Note that $\mathbb{S}_{i,k}$ may include interaction terms with baseline variables. Here, $\hat{\beta}$ can be computed by maximum likelihood or finding the roots of the score equation. 

After the pooled logistic regression model is fit, the predicted probability of survival up to interval $k$ is computed from the parameter estimates of the pooled logistic regression model. First, the probability of $Y_{i,k} = 1$ conditional on survival to the start of interval $k$ is predicted from the model. These predictions correspond to the discrete time hazard for person-period $i,k$. The discrete time hazard can be transformed into the probability of survival through interval $k$ by taking the cumulative product of the complement of the discrete time hazards through interval $k$, i.e., $\prod_{j=1}^{k} \Pr(Y_{i,j} = 0 \mid A_i, W_i, T_i^* > s_{j-1}; \hat{\beta})$. In the case of g-computation for estimating causal effects, the predicted discrete time hazards are generated when $A$ is set to different values (e.g., set to $1$ for all units), $\mu_a(s_k, W_i; \hat{\beta}) = 1 - \prod_{j=1}^{k} \Pr(Y_{i,j} = 0 \mid A_i = a, W_i, T_i^* > s_{j-1}; \hat{\beta})$ and the estimator in Equation \ref{gcomp_estr} can be applied. However, $A$ can also remain as observed to estimate the `natural course' \cite{rudolphRoleNaturalCourse2021}.

\subsection{Functional Form for Time}

As implied by Equation \ref{PLRM}, the analyst must specify some functional form for time in the design matrix $\mathbb{S}$. The chosen functional form for time imposes constraints on how the discrete time hazard can vary across time. Modifying the function form allows for constraints on changes in the hazard function that are analogous to constraints imposed by other popular continuous time survival analysis models. If time is modeled using $K$ disjoint indicators (i.e., there is a unique parameter in $\beta_S$ for each time interval), then no constraints are placed on how the discrete time hazard changes over time. This modeling assumption bears similarity to the assumptions of a Cox proportional hazards model, a semiparametric model that places no constraints on the baseline hazard function. Specifically, $\exp(\beta_X)$ approximate the hazards ratio from a Cox proportional hazards model when events are rare within each time interval \cite{youngRelationThreeClasses2010, murrayCausalSurvivalAnalysis2021a}. Alternatively, one can have $\mathbb{S}$ include only a constant for all units (i.e., only an intercept). Such a model assumes the discrete time hazard is the same for each interval. This constraint on the discrete time hazard is analogous to the constraint on continuous time hazards in an exponential model. 
Between these two extreme assumptions on the shape of the discrete time hazard function, one could consider constraints analogous to those imposed by other popular parametric continuous time survival models. Linear and log-linear constraints on the discrete time hazard are analogous to the constraints imposed by Gompertz and Weibull models, respectively \cite[pg.~242]{allisonSurvivalAnalysisUsing2010}. Another option has been to use splines to model time \cite{hernanHazardsHazardRatios2010, wanisEffectOpioidEpidemic2019, murrayCausalSurvivalAnalysis2021a}, as splines impose less restrictive functional form assumptions on how the discrete time hazard changes over time but does not involve estimating as many parameters as using disjoint indicators. Overall, one can view pooled logistic regression models as a bridge between a semiparametric Cox model and parametric survival models in the discrete time setting. Specifically, one can add (or remove) constraints on the functional form for time to restrict (or relax) how the hazard function can change over time. 

\section{Estimating Equations}

Now we show how a pooled logistic regression model can be expressed with estimating equations. An estimating equation is defined as $E[\psi(O_i; \theta)] = 0$, where $\psi(O_i; \theta)$ is a $b$-dimensional estimating function, and $\theta$ is a $b$-dimensional vector of parameters. Hereafter, $b$ is considered to be finite. The corresponding estimator, $\hat{\theta}$ (often referred to as a Z-estimator or M-estimator \cite{stefanskiCalculusMEstimation2002, boosMEstimationEstimatingEquations2013, vaartMandZEstimators1998, kosorokZEstimators2008}), is the solution to the vector equation $n^{-1} \sum_{i=1}^{n} \psi(O_i; \theta) = 0$. Under suitable regularity conditions \cite{boosMEstimationEstimatingEquations2013}, the corresponding estimator is consistent and asymptotically normal for $\theta_0$ if the estimating equations are unbiased at the true value of the parameters (i.e., $E[\psi(O_i; \theta_0)] = 0$ where $\theta_0$ denotes the true value) \cite{boosMEstimationEstimatingEquations2013}. Further the asymptotic variance of $\hat{\theta}$ can be consistently estimated by the empirical sandwich variance estimator $V(\hat{\theta}) = B(\hat{\theta})^{-1} M(\hat{\theta}) \left[B(\hat{\theta})^{-1}\right]^T$, where $B(\theta) = n^{-1} \sum_{i=1}^{n} \nabla_{\theta} \psi(O_i; \theta)$ with $\nabla_{\theta}$ denoting the gradient of the estimating function with respect to $\theta$ and $M(\theta) = n^{-1} \sum_{i=1}^{n} \psi(O_i; \theta) \psi(O_i; \theta)^T$. The empirical sandwich variance estimator can be used to \edit{directly } construct Wald-type CIs \edit{for a single parameter in $\theta$ } or used to construct Wald-type CBs using the sup-t method \edit{for a subset of parameters in $\theta$ } \cite{zivichConfidenceRegionsMultiple2026}. \edit{Importantly, the empirical sandwich variance estimator is consistent for $\hat{\theta}$ regardless of whether the point estimator is unbiased for $\theta_0$. } For further reading on estimating equations, see the following references \cite{stefanskiCalculusMEstimation2002, boosMEstimationEstimatingEquations2013, rossMestimationCommonEpidemiological2024}.

\edit{
	Estimating equations have several features relevant to our setting. First, separate estimating functions for different parameters can be stacked together. These stacked estimating functions can then be used to simultaneously estimate multiple parameters \cite{raymondj.carrollMeasurementErrorNonlinear2006}. This stacking is particularly useful when the estimating function for one parameter contains parameters that appear in other estimating functions, as in settings with nuisance parameters like with g-computation. Stacking estimating functions also means that uncertainty in estimating parameters is appropriately propagated when an estimating function depends on the parameters of another. This feature means that the uncertainty introduced by estimation of nuisance parameters is incorporated into estimation of the asymptotic variance of the target parameter. Second, it is straightforward to estimate the variance of smooth transformations of parameters \cite{stefanskiCalculusMEstimation2002}. The estimating function for the transformed parameter $\theta_g$ is simply $g(\theta) - \theta_g = 0$, where $g$ is the function applying the transformation to $\theta$. Note that this estimating function does not depend on $O_i$. In this way, the delta method can be implemented through estimating equations \cite{stefanskiCalculusMEstimation2002}. Finally, the empirical sandwich variance estimator allows for units to provide multiple contributions, i.e., clustered observations. Aggregating contributions from units can be generalized through use of the working covariance matrix. Importantly, inference is robust to misspecification of the working covariance matrix \cite{boosMEstimationEstimatingEquations2013}. 
}

\subsection{Pooled Logistic Regression Models as Estimating Equations}

Now consider how a pooled logistic regression model can be \edit{expressed as an estimating function. The formulation here allows for two computational improvements: (1) avoiding the need to create a long data set and (2) variance estimation for the target parameter through the empirical sandwich variance estimator. } As previously indicated, the parameters of the pooled logistic regression model can be estimated by finding the values of $\hat{\beta}$ where the score equation is equal to zero. For the score equation, each person-period contributes once and these contributions are added together. By the commutative property of addition, each of these contributions can be re-ordered. Rather than sum across all rows of a long data set, all the contributions over time for a single unit can instead be summed together. Unit-specific contributions can then be added together. This re-ordering of summations gives the following estimating function for $\beta$ for unit $i$
\begin{equation}
	\psi(O_i; \beta) = \sum_{k \in \mathcal{K}} \left\{ I(T_i^* > s_{k-1}) \left( Y_{i,k} -  \hat{Y}_{i,k} \right) 
	\begin{bmatrix}
		\mathbb{X}_i^T \\
		\mathbb{S}_{i,k}^T \\
	\end{bmatrix}
	\right\}
	\label{EE1}
\end{equation}
where $\hat{Y}_{i,k} = \text{expit}(\mathbb{X}_i \beta_X + \mathbb{S}_{i,k} \beta_S)$. The $I(T_i^* > s_{k-1})$ term restricts contributions to those in the risk set at the start of interval $k-1$. Alternatively, one may have weights associated with each person-period (e.g., sampling weights in cases of probabilistic surveys). The previous estimating function can be replaced by the score function for a weighted logistic regression model to incorporate unit- and time-specific weights,
\begin{equation}
	\psi_{w}(O_i; \beta) = \sum_{k \in \mathcal{K}} \left\{ I(T_i^* > s_{k-1}) \left( Y_{i,k} - \hat{Y}_{i,k} \right) 
	\begin{bmatrix}
		\mathbb{X}_i^T \\
		\mathbb{S}_{i,k}^T \\
	\end{bmatrix}
	\omega_{i,k} \right\}
	\label{EE2}
\end{equation}
where $\omega_{i,k}$ is the weight for unit $i$ in the $k$ interval. \edit{When $\omega_{i,k}$ is estimated rather than fixed, the corresponding estimator for the weights should also be added, or stacked, into the system of estimating equations for valid variance estimation.}

By maximum likelihood theory, it follows that the estimating equations are unbiased \cite{demonteAssessingCOVID19Vaccine2024}. Therefore, $\hat{\beta}$ is consistent and asymptotically normal following estimating equation theory for finite parameter vectors under suitable regularity conditions \cite{stefanskiCalculusMEstimation2002}. \edit{Notice that re-ordering the summation leads to a difference between the model-based and empirical sandwich variance estimators for $\beta$. The summation over $\mathcal{K}$ in \eqref{EE1} and \eqref{EE2} accumulates the person-period-specific contributions into person-specific contributions. Functionally, this formulation of the pooled logistic regression model is equivalent to using Generalized Estimating Equations (GEE) with an independent working covariance matrix where `person' is designated as the cluster.}

To convert the unit-specific discrete time hazards into the marginal risk at \edit{$m \in \mathcal{K}$ } with $A=a$, the following estimating function can be used
\begin{equation}
	\psi_r(O_i; \beta, \gamma^a_m) = \left\{ 1 - \prod_{k \in \{1, ..., m\}} \left[ 1 - \hat{Y}^a_{i,k} \right] \right\} - \gamma^a_m
	\label{EE3}
\end{equation}
where \edit{$\hat{Y}^a_{i,k}$ } is the predicted discrete time hazard \edit{at $s_k$ } when $A$ is set to $a$ in $\mathbb{X}$, and \edit{$\gamma^a_m \equiv F^a(s_m)$}. Other quantities (e.g., survival, cumulative hazard) can be obtained through transformations of $\gamma^a_m$. \edit{The causal risk difference can then be estimated by $\psi_{rd}(O_i; \gamma^1_m, \gamma^0_m, \rho_m) = \gamma^1_m - \gamma^0_m  - \rho_m$, where $\rho_m \equiv F^1(s_m) - F^0(s_m)$. } For g-computation, $\psi_{rd}$ is unbiased following the assumptions for \ref{id_result} and correct specification of the pooled logistic regression model. 
\edit{Stacking the estimating functions for the g-computation estimator allows for variance estimation of $\hat{\rho}_m$ with person-specific contributions through the stacked sandwich estimator for the nuisance parameters, risks, and risk difference. }

\subsection{Solving Estimating Equations}

To estimate the parameters of \eqref{EE1}-\eqref{EE3}, one can use a root-finding algorithm (e.g., Levenberg-Marquardt \cite{levenbergMethodSolutionCertain1944}, Powell hybrid \cite{powellEfficientMethodFinding1964}, secant \cite{ehiwarioComparativeStudyBisection2014}) to find an approximate zero of the estimating equations. After computing $\hat{\theta}$, $B(\hat{\theta})$ can be \edit{manually derived symbolically (Appendix A) or computed for particular values of $\theta$ via algorithmic procedures (e.g., numerical approximation, automatic differentation). } Matrix operations to compute $M(\hat{\theta})$ and $V(\hat{\theta})$ are also available in standard statistical analysis software. While these computations can be implemented by-hand, these procedures are automated for user-specified estimating equations in the open-source libraries \texttt{delicatessen} for Python and \texttt{geex} for R \cite{zivichDelicatessenMEstimationPython2022, saulCalculusMEstimationGeex2020}.

While the estimating function in \eqref{EE1} can be implemented through a for-loop over $\mathcal{K}$, this implementation can be relatively slow for interpreted languages (e.g., Python, R). Any computational burdens in evaluating the estimating functions is a particular issue, as the estimating function must be called during each iteration of the root-finding procedure. Therefore, we describe a vectorization of the estimating functions. To help describe this vectorization, additional notation is introduced. Let bold fonts denote \edit{column } vectors \edit{with subscripts denoting the dimension } (e.g., $\mathbf{V}_r$), and calligraphic fonts denote matrices with subscripts denoting their dimensions (e.g., $\mathcal{M}_{r \times c}$ is a matrix with $r$ rows and $c$ columns). For two matrices $\mathcal{M}$ and $\mathcal{N}$, let $\mathcal{M} \cdot \mathcal{N}$ denote the usual dot product for conforming matrices and $\mathcal{M} \odot \mathcal{N}$ denote the Hadamard, or element-wise, product between two matrices or vectors of the same dimension. \edit{In a minor abuse of notation, the $\oplus$ operator with a matrix and vector is used as shorthand for $\mathcal{M}_{r \times c} + (\mathbf{V}_r \otimes 1_{1\times c})$, where $\otimes$ is the Kronecker product. That is, }
\begin{equation*}
	\mathcal{M} \oplus \mathbf{V} =
	\begin{bmatrix}
		m_{11} & \dots  & m_{1c} \\
		\vdots & \ddots & \vdots \\
		m_{r1} & \dots  & m_{rc} 
	\end{bmatrix}
	\oplus 
	\begin{bmatrix}
		v_{1}  \\
		\vdots \\ 
		v_{r} 
	\end{bmatrix}
	= 
	\begin{bmatrix}
		m_{11} + v_{1}  & \dots  & m_{1c} + v_{1} \\
		\vdots & \ddots & \vdots \\
		m_{r1} + v_{r}  & \dots  & m_{rc} + v_{r} 
	\end{bmatrix}
\end{equation*}
\edit{Similarly, the $\odot$ operator between a matrix and vector is used as shorthand for }
\begin{equation*}
	\mathcal{M} \odot \mathbf{V} =
	\begin{bmatrix}
		m_{11} \times v_{1}  & \dots  & m_{1c} \times v_{1} \\
		\vdots & \ddots & \vdots \\
		m_{r1} \times v_{r}  & \dots  & m_{rc} \times v_{r} 
	\end{bmatrix}.
\end{equation*}
Finally, $\text{expit}(\cdot)$ of a matrix or vector is defined as the element-wise application of the $\text{expit}$ function.

To begin, a vectorization for a parametric specification of the functional form for time is described. Let $\mathbf{1}_{r \times c}$ denote a $r$ by $c$ matrix of ones,
\begin{equation*}
	\mathbf{\Delta}_{K \times n} = 
	\begin{bmatrix}
		\Delta_1 & \dots  & \Delta_n \\
		\vdots   & \ddots & \vdots \\
		\Delta_1 & \dots  & \Delta_n \\		
	\end{bmatrix}
\end{equation*} 
and
\begin{equation*}
	\mathcal{X}_{K \times n} = 
	\begin{bmatrix}
		(\mathbb{X}_{p \times n} \cdot \beta_X)^T \\
		\vdots  \\
		(\mathbb{X}_{p \times n} \cdot \beta_X)^T
	\end{bmatrix}
\end{equation*}
where \edit{$\mathbb{X}_{p \times n}$ is the complete design matrix for the $n$ observations and $p$ terms}, $(\mathbb{X}_{p \times n} \cdot \beta_X)^T$ is a \edit{row } vector of the $n$ log-odds predictions (when $\beta$ is replaced by $\hat{\beta}$). Let $\mathcal{S}_{K \times q}$ denote the design matrix for each $s_k$, where the rows are in order from $1$ to $K$ and the columns correspond to the generated terms (e.g., the $k$\textsuperscript{th} row for a linear model of time would be $[1 \; k]$). Let $\mathcal{R}_{K \times n}$ be a matrix of indicator terms designating whether a unit was present in the risk set, where the rows corresponds to ascending discrete times up to $K$ and columns correspond to observations (e.g., elements in each column are equal to one until after the interval a unit had the event or was censored, and is zero thereafter). Let $\mathcal{R}^*_{K \times n}$ denote a matrix like $\mathcal{R}_{K \times n}$ but instead only indicates the final time under observation (e.g., the $i$\textsuperscript{th} column of $\mathcal{R}^*$ is zero except for the $k$ row when unit $i$ had the event or was censored). Let $\mathcal{Y}_{K \times n} = \mathbf{\Delta} \odot \mathcal{R}^*$ indicate the observed events at their corresponding times and $\hat{\mathcal{Y}}_{K \times n} = \text{expit}\left( \mathcal{X}_{K \times n} \oplus \mathcal{S}_{K \times q} \cdot \edit{\beta_S^T} \right)$ be a matrix of the discrete time hazards.

Therefore, a $n \times p$ matrix for the $p$ parameters in $\beta_X$ can be obtained by $\left(\mathbf{1}_{1 \times K} \cdot \mathcal{P}_{K \times n} \right)^T \odot \mathbb{X}_{p \times n}$, where $\mathcal{P} = (\mathcal{Y} - \hat{\mathcal{Y}}) \odot \mathcal{R}$ is a units contribution to the estimating equation for each time interval. The $q \times n$ matrix for the $q$ parameters in $\beta_S$ can be obtained via $\mathcal{S}^T \cdot \mathcal{P}$. \edit{With disjoint indicators for time, the estimating function matrix for $\beta_S$ is directly obtained via $\mathcal{P}$, rather than $\mathcal{S}^T \cdot \mathcal{P}$}. These matrices can then be stacked together to obtain the full $p+q$ by $n$ matrix of estimating functions for $\beta$. A vectorization of the predicted survival for each unit can be obtained by taking the column-wise cumulative product of the complement of $\hat{\mathcal{Y}}$. From there, one can select the row(s) that correspond to discrete time(s) of interest. Appendix B provides an illustrative example of these computations.

In the case of disjoint indicators for time, the preceding vectorization can be \edit{further improved. Specifically, $\mathbb{S}$ only needs to include disjoint indicator terms for unique event times \cite{zivichImprovedPooledLogistic2026}. } Let $K^*$ denote the number of unique event times \edit{and $\mathcal{K}^*$ denote the index set for the intervals that contain an event}. 
\edit{To see why intervals outside of $\mathcal{K}^*$ can safely be ignored, notice that the estimated discrete-time hazard must be zero as there are no events in those intervals and information is not borrowed from other intervals. Therefore, coefficients for indicators of time intervals without events should be $-\infty$ as $\text{expit}(-\infty) = 0$. With the standard implementation, this behavior is approximated through floating point errors and small negative values (i.e., $\text{expit}(-30) < 1 \times 10^{-13}$). Rather than approximate zeroes, these intervals can simply be skipped as they do not contribute additional information for the risk or hazard. For the estimating equations, this modification means that fewer coefficients need to be estimated and } the previous matrices shrink their corresponding dimension from $K$ to $K^*$. \edit{This same approach can also be used to improve computation } with the standard implementation and disjoint indicators by restricting the long data set to rows corresponding to unique event times \cite{zivichImprovedPooledLogistic2026}, \edit{but is not considered further here.}

Generally, the vectorization of the estimating equations is expected to use less memory than the standard implementation. For the standard implementation, a total of $(p + q + 1) K n$ elements are stored. The vectorization of the estimating equation consists of $n + 5Kn + Kq$ elements. So, less elements exist whenever the following inequality holds, $n + 5Kn + Kq < Kn + Knp + Knq$. Pairing $n < Kn$, $Kq < Knq$, and $5Kn < Knp$; one can see that the vectorization is expected to be smaller when either $K$ or $n$ becomes large. In practice, $K$ or $n$ is expected to be large. The reduction in elements is expected to be even greater for the disjoint indicator implementation when there are few unique event times relative to the number of discrete time points, as the inequality is instead $n + 5K^* n + K^* q < (p + q + 1) K n$. Regardless, the vectorization still requires more memory than a for-loop implementation. A for-loop implementation instead only ever needs to store $n(p+q+3)$ elements in actively memory (i.e., $\mathbb{S}_{i,k}$ can be overwritten for each $k \in \mathcal{K}$). Therefore, one can opt for the for-loop implementation when there are constraints on the available memory at the cost of run-time. This flexibility of implementation based on available computational resources is not available for the standard approach.

\section{Examples}

To demonstrate the \edit{proposed estimating equations for g-computation with } pooled logistic regression, two examples using publicly available data are provided. In addition to the \edit{causal risk difference } reported at $\tau$ with 95\% CIs, the risks and risk differences are presented graphically over the duration of follow-up, along with their sup-t 95\% CBs. The risk difference function and CBs are displayed using twister plots \cite{zivichTWISTERPLOTSTIMETOEVENT2021}.

In each example, results are reported for the standard and estimating equation implementations. To compare performance, run-times are reported \edit{for the complete point and variance estimation process } using the \edit{median } across 5 repetitions of the same implementation in Python 3.13.7. Performance was measured using a laptop running Windows 11 Enterprise on the `best performance' power setting with an AMD Ryzen\texttrademark AI Pro 350 processor (8-core, 2.00 GHz) and 64GB of RAM. Variance estimation for 95\% CIs the standard implementation was accomplished using the fractional-random-weight bootstrap with 1000 iterations. Bootstrap run-time results were separately reported for all bootstraps in sequence and up to 7 in parallel. For the estimating equations, the vectorized implementation was used.

\subsection{Example 1}

The first example comes from a placebo-controlled trial on disease-free survival among bladder cancer patients \cite{byarComparisonsPlaceboPyridoxine1977}, with data obtained from Collett \cite{collettModellingSurvivalData2015}. The available data consists of 86 patients who received either placebo or the chemotherapeutic thiotepa following removal of superficial bladder tumors. Time was measured in months up to $\tau = 59$. Other baseline variables collected included the initial number of tumors and the diameter (in centimeters) of the largest initial tumor. Here, the parameter of interest was the \edit{causal risk difference } comparing novel to standard treatment on \edit{the composite event of cancer recurrence or death } at 59 months. While treatment was randomized, adjustments were made for both number of tumors and tumor diameter. Both variables were modeled as linear relationships for simplicity. Time was modeled separately using disjoint indicators and restricted quadratic splines with knots at 10, 20, 30, and 40 months.

\begin{figure}
	\centering
	\caption{Estimated risks and risk differences of \edit{cancer recurrence or death } among bladder cancer patients comparing thiotepa versus placebo from example 1 with time modeled using (A) disjoint indicators and (B) restricted quadratic splines}
	\includegraphics[width=0.85\linewidth]{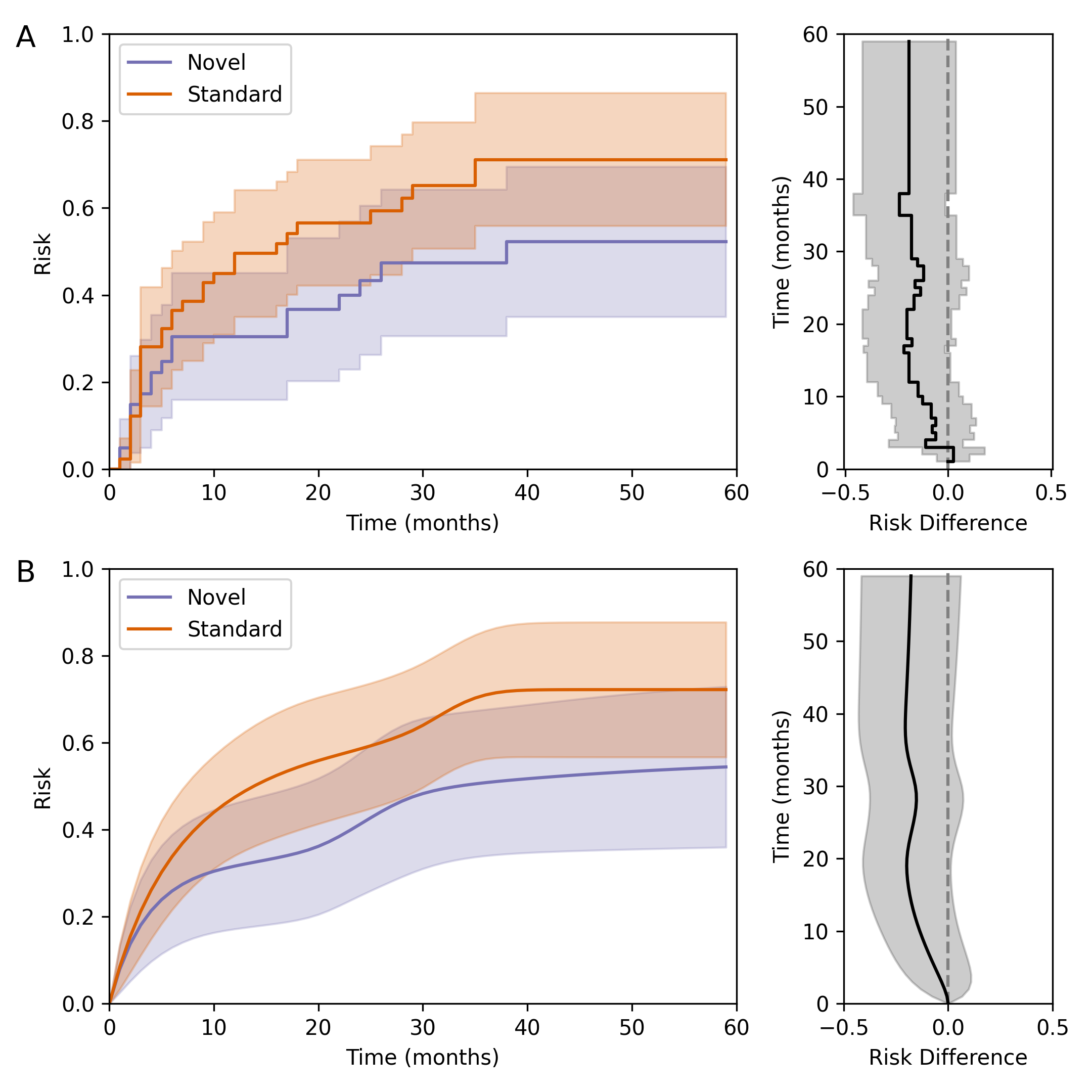} 
	\floatfoot{Shaded regions represent 95\% confidence bands estimated using the sup-t method and the empirical sandwich variance estimator.}
	\label{Fig1}
\end{figure}

Estimated risks and corresponding 95\% CBs are presented in Figure \ref{Fig1}. At 59 months, the estimated \edit{causal risk difference } comparing thiotepa to placebo was $-0.19$ (95\% CI: -0.42, 0.04, Table \ref{Table1}). The CIs were similar \edit{between variance estimators}. Regarding run-times, there were notable differences. For either functional form of time, the estimating equation implementation took less than one second. With splines, the run-time was slightly longer, most likely due to the need to evaluate the estimating functions at each time interval unlike the disjoint indicator implementation. With the standard implementation, run-times were more than a minute when modeling time with disjoint indicators. The standard implementation run-time benefited greatly when modeling time using splines. 

\begin{table}[]
	\caption{Results for Example 1}
	\begin{tabular}{lllccc}
		\hline
		&           &                              & RD    & 95\% CI     & Run-time\textsuperscript{*} \\ \cline{4-6} 
		\multicolumn{3}{l}{Disjoint Indicator}          &       &             &          \\
		& \multicolumn{2}{l}{Estimating Equations} & -0.19 & -0.42, 0.04 & 0.2      \\
		& \multicolumn{2}{l}{Standard}             &       &             &          \\
		&           & Bootstrap in Sequence        & -0.19 & -0.42, 0.04 & 709.8    \\
		&           & Bootstrap in Parallel        & -0.19 & -0.42, 0.04 & 106.0    \\
		\multicolumn{3}{l}{Splines}                     &       &             &          \\
		& \multicolumn{2}{l}{Estimating Equations} & -0.18 & -0.42, 0.06 & 0.7      \\
		& \multicolumn{2}{l}{Standard}             &       &             &          \\
		&           & Bootstrap in Sequence        & -0.18 & -0.42, 0.06 & 22.9     \\
		&           & Bootstrap in Parallel        & -0.18 & -0.42, 0.06 & 6.7     \\ \hline
	\end{tabular}
	\floatfoot{RD: risk difference at $\tau$, CI: Wald-type confidence intervals. \\
		\textsuperscript{*} Run-times are the median of five runs and reported in seconds.}
	\label{Table1}
\end{table}

\subsection{Example 2}

The second example comes from Lau et al. (2009) \cite{lauCompetingRiskRegression2009}, which analyzed data on 1164 women with HIV from the Women's Interagency HIV Study (WIHS) \cite{barkanWomensInteragencyHIV1998}. Baseline covariates included history of injection drug use (yes; no), race (Black; White), nadir CD4 T cell count (in 100 cells / mm\textsuperscript{3}), and age (in years). The event of interest was the composite event of AIDS or death. The parameter of interest was the \edit{causal risk difference } on the composite outcome comparing history of injection drug use to no history at 3653 days (10 years). Those who started antiretroviral treatment were right censored, so the effect is defined for those who did not start treatment. Confounding variables included race, age, and nadir CD4. Nadir CD4 and age were modeled using restricted quadratic splines with knots at $2.1,3.5,5.2$ and $25, 35, 50$, respectively. Note that this example should be viewed as \edit{a computational illustration } and not a rigorous causal analysis. In particular, \edit{attempts to modify } historic injection drug use induces selection bias, as \edit{a true intervention } on past drug use would likely prevent some of the HIV infections from occurring, which \edit{subsequently } modifies the population of interest. This issue is similar those arising in studying causal effects of vaccines on post-infection outcomes \cite{hudgensCausalVaccineEffects2006}.

To emphasize the challenge faced by the standard implementation, note that a long data set consists of up to $1164 \times 3653 = 4,252,092$ rows. When modeling time using disjoint indicators, there are 3652 parameters for time alone, most of which will be small negative numbers (e.g., $<-20$) as no events occur in those intervals \cite{thompsonTreatmentGroupedObservations1977}. When attempting to use the standard implementation with disjoint indicators, there was not enough memory available to fit this model. Using splines, a pooled logistic model could be fit but the bootstrap could not be run in parallel due to memory. With splines, fitting this model could take several \textit{hours}. A common strategy to ease the computational burden of the standard implementation is to discretize time into larger units. Here, we coarsened time from days to months. The standard implementation was able to be applied in most settings (Table \ref{Table2}), but modeling time with disjoint indicators still took hours to estimate the variance. 

\begin{figure}
	\centering
	\caption{Risks and risk differences of \edit{AIDS or death } among women with HIV comparing history of injection drug use versus none in example two with time modeled using (A) disjoint indicators and (B) restricted quadratic splines}
	\includegraphics[width=0.85\linewidth]{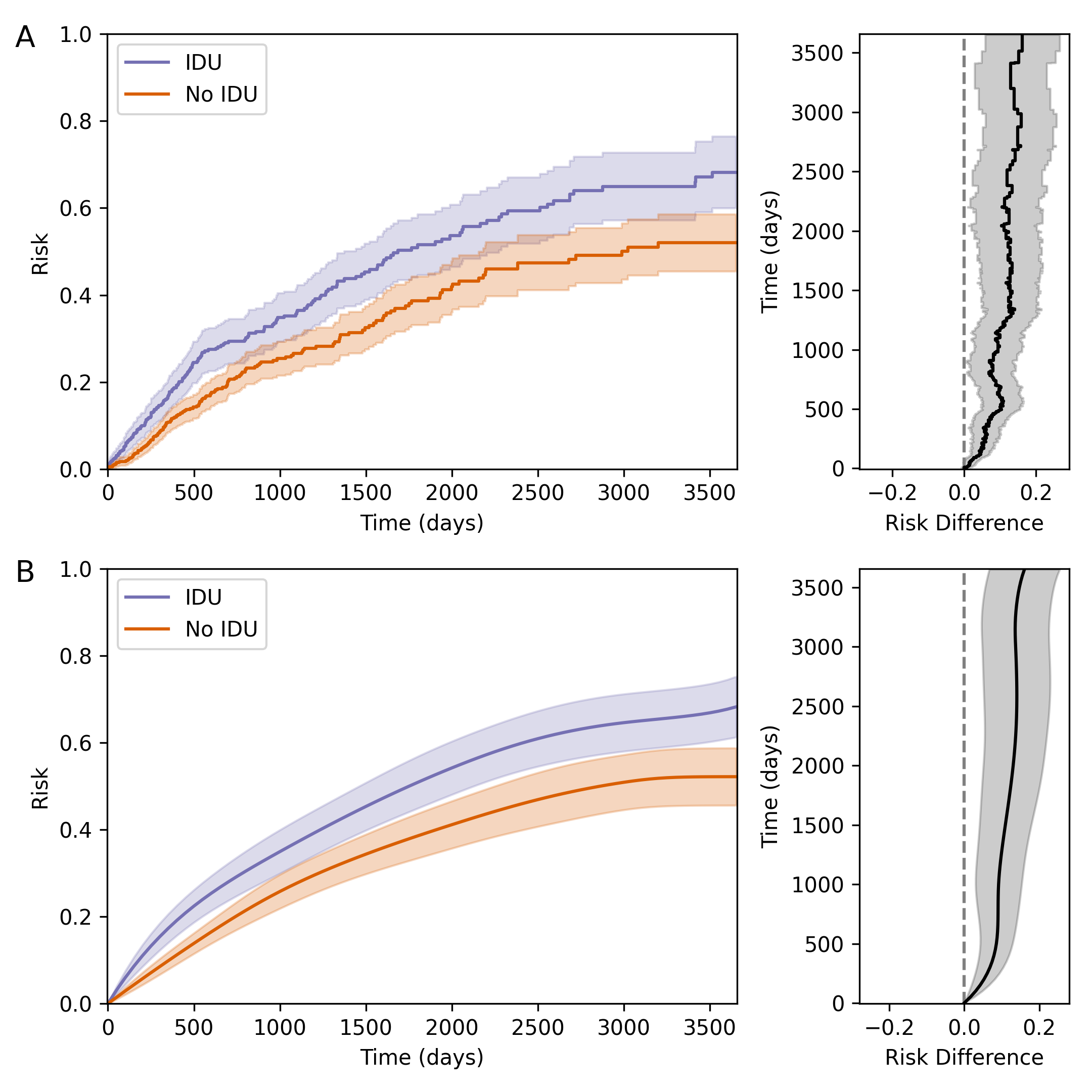}
	\floatfoot{Shaded regions represent 95\% confidence bands estimated using the sup-t method and the empirical sandwich variance estimator.}
	\label{Fig2}
\end{figure}

Here, the estimating equation implementation provided results for both functional forms of time on both time scales (Table \ref{Table2}). Those estimated risks \edit{over time } and the corresponding 95\% CBs from the estimating equation implementation are presented in Figure \ref{Fig2}. When using time in days, the estimating equations took no more than two minutes to run. When switching to the months scale, run-times were in the seconds. The 95\% CIs were comparable between bootstrapping and the empirical sandwich variance estimator when both could be applied.

\begin{table}[]
	\caption{Results for Example 2}
	\begin{tabular}{lllccccccc}
		\hline
		&           &                              & \multicolumn{3}{c}{Time in Days} &  & \multicolumn{3}{c}{Time in Months} \\ \cline{4-6} \cline{8-10} 
		&           &                              & RD     & 95\% CI      & Run-time\textsuperscript{*}  &  & RD     & 95\% CI       & Run-time\textsuperscript{*}   \\ \cline{4-6} \cline{8-10} 
		\multicolumn{3}{l}{Disjoint Indicator}          &        &             &           &  &        &              &            \\
		& \multicolumn{2}{l}{Estimating Equations} & 0.16   & 0.06, 0.27  & 55.0      &  & 0.16   & 0.06, 0.27   & 5.0        \\
		& \multicolumn{2}{l}{Standard}             &        &             &           &  &        &              &            \\
		&           & Bootstrap in Sequence        & -      & -           & -         &  & 0.16    & 0.07, 0.26  & $>$9000.0          \\
		&           & Bootstrap in Parallel        & -      & -           & -         &  & -    & -  & -          \\
		\multicolumn{3}{l}{Splines}                     &        &             &           &  &        &              &            \\
		& \multicolumn{2}{l}{Estimating Equations} & 0.16   & 0.07, 0.25  & 113.4     &  & 0.16   & 0.06, 0.27   & 2.7       \\
		& \multicolumn{2}{l}{Standard}             &        &             &           &  &        &              &            \\
		&           & Bootstrap in Sequence  & 0.16   & 0.07, 0.26      & $>$15,000.0    &  & 0.16   & 0.07, 0.26   & 439.5      \\
		&           & Bootstrap in Parallel        & -      & -           & -         &  & 0.16   & 0.07, 0.26   & 171.4      \\ \hline
	\end{tabular}
	\floatfoot{RD: causal risk difference, CI: Wald-type confidence intervals. \\
		\textsuperscript{*} Run-times are the median of five runs and reported in seconds. Note that the standard implementation run-time results for those taking at least an hour are only based on a single run due to their long computation time.}
	\label{Table2}
\end{table}

\section{Simulations}

To illustrate the performance of the empirical sandwich variance estimator for inference with g-computation, a Monte Carlo simulation study was conducted. The parameters of interest were the \edit{causal risk differences } at {$t \in \{15, 30\}$. The observed data was generated from the following mechanisms. Let $\lceil \cdot \rceil$ denote the ceiling function. A continuous baseline confounder was simulated via $ W_i \sim \text{Uniform}(-1,1) $. Treatment was then assigned as $ A_i \sim \text{Bernoulli}\left[ \text{expit}(-1.5 \times W_i) \right] $. The potential event times under $A=1$ and $A=0$ were simulated from Weibull distributions as $ T_i^1 \sim \left\lceil (50 + 5 \times W_i + 15) \times \left[ \ln(- \text{Uniform}(0,1)) \right]^{1/0.75} \right\rceil $ and $ T_i^0 \sim \left\lceil (50 + 5 \times W_i + 0) \times \left[ \ln(- \text{Uniform}(0,1)) \right]^{1/1.5} \right\rceil $, respectively. The event time under the observed treatment was then $T_i = T_i^1 A_i + T_i^0 (1-A_i)$. Non-informative censoring times were simulated from $C_i = \lceil 38 \times \ln(-\text{Uniform}(0,1)) \rceil$. Finally, the observed time was $T_i^* = \min(T_i, C_i)$ and $\Delta_i =  I(T_i = T_i^*)$.

For estimation of the parameter of interest, variations on specifications of the pooled logistic regression model were considered. Five specifications were compared. First, $\mathbb{S}$ was composed of only an intercept term. Second, $\mathbb{S}$ included both an intercept term and linear term for time. Third, $\mathbb{S}$ included an intercept and the natural log of time. Fourth, $\mathbb{S}$ included an intercept term, linear term, and restricted quadratic splines with knots at 5, 10, 15, 20, 25 were used. Finally, $\mathbb{S}$ was composed of disjoint indicator terms, imposing no functional form assumptions on time. In all pooled logistic models, $W$ was modeled using a restricted cubic spline with knots at -0.8, 0, 0.8. Each model was fit separately by values of $A$. Here, the first two models are expected to be biased at some times due to misspecification of the functional form for time, while the remainder are expected to be unbiased. Further, the pooled logistic model with a log-linear specification for time is expected to have the smallest variance as it is
\edit{approximately correctly specified } with the fewest additional parameters estimated.

To evaluate the performance of each estimator and the empirical sandwich variance estimator, the following metrics were considered at each $t$ of interest: bias, empirical standard error (ESE), average standard error (ASE), standard error ratio (SER), and 95\% CI coverage \cite{morrisUsingSimulationStudies2019}. Bias was defined as the estimated \edit{risk difference } minus the true \edit{causal risk difference } (approximated by simulating the potential outcomes for 10 million observations under the data generating process described above). ESE was estimated by taking the standard deviation of the point estimates across simulations. ASE was estimated by taking the average of the standard error estimates from the empirical sandwich variance estimator. SER was defined as the ASE divided by the ESE. An SER near one indicates good performance of the variance estimator, with values smaller than one indicating under-estimation. Finally, CI coverage was estimated by the proportion of 95\% CIs that contained the true \edit{causal risk difference at $t$}. These metrics were evaluated across 5000 iterations at four different sample sizes: $n \in \{250, 500, 1000, 2000\}$. Here, performance of the bootstrap was not evaluated as the empirical sandwich variance estimator was of interest and bootstrapping would dramatically increase the overall runtime of the simulations. Simulations were conducted with Python 3.9.4 (Python Software Foundation, Beaverton, OR, USA) and the following packages: \texttt{NumPy}, \texttt{SciPy}, \texttt{delicatessen}, and \texttt{pandas}. Code to replicate the simulations is provided at \url{https://github.com/pzivich/publications-code}.

\subsection{Results}

For the pooled logistic models that \edit{were approximately correctly specified } for time (i.e., log-linear, splines, disjoint indicators), bias was near zero, SER was near 1, and CI coverage was near nominal for all sample sizes \edit{and times } (Table \ref{TableSimResults}). These results agree with the theoretical expectations. Here, modeling time more flexibly than necessary led to an increased ESE, as expected. This result was most apparent for the smaller sample sizes and at the earlier time points. When modeling time with an inappropriate functional form (i.e., intercept-only, linear), bias occurred for at least one of the time points with correspondingly poor CI coverage. This was most notable for the intercept-only model, which as biased at both time points considered. The linear model for time only had noticeable bias at $t=15$. Additional simulation results for bias and coverage are provided in the Appendix C for $t \in \{5, 10, 15, 20, 25, 30\}$. Altogether, these simulation results support the theoretical expectations for the performance of the empirical sandwich variance estimator in this setting and highlight the importance of modeling time flexibly when the functional form for the discrete time hazards is unknown (which is often the case in practice).

\begin{table}[H]
	\caption{Simulation results for estimation of the causal risk difference using g-computation with pooled logistic regression and estimating equations across sample sizes}
	\centering	
	\begin{tabular}{lccccccccc}
		\hline
		& \multicolumn{4}{c}{$t=15$}                                                                               &  & \multicolumn{4}{c}{$t=30$}                                                                               \\ \cline{2-5} \cline{7-10} 
		& \multicolumn{1}{c}{Bias} & \multicolumn{1}{c}{ESE} & \multicolumn{1}{c}{SER} & \multicolumn{1}{c}{Cover} &  & \multicolumn{1}{c}{Bias} & \multicolumn{1}{c}{ESE} & \multicolumn{1}{c}{SER} & \multicolumn{1}{c}{Cover} \\ \cline{2-5} \cline{7-10} 
		$n=250$        & \multicolumn{1}{c}{}     & \multicolumn{1}{c}{}    & \multicolumn{1}{c}{}    & \multicolumn{1}{c}{}      &  & \multicolumn{1}{c}{}     & \multicolumn{1}{c}{}    & \multicolumn{1}{c}{}    & \multicolumn{1}{c}{}      \\
		Intercept-only & -0.047                   & 0.053                   & 0.98                    & 0.84                      &  & 0.069                    & 0.079                   & 0.98                    & 0.84                      \\
		Linear         & 0.016                    & 0.057                   & 0.98                    & 0.93                      &  & -0.007                   & 0.085                   & 0.98                    & 0.94                      \\
		Log-linear     & 0.005                    & 0.055                   & 0.97                    & 0.94                      &  & -0.005                   & 0.084                   & 0.99                    & 0.94                      \\
		Spline         & 0.000                    & 0.060                   & 1.00                    & 0.94                      &  & -0.001                   & 0.085                   & 0.98                    & 0.94                      \\
		Disjoint       & -0.001                   & 0.062                   & 0.99                    & 0.95                      &  & -0.001                   & 0.085                   & 0.98                    & 0.94                      \\
		$n=500$        & \multicolumn{1}{c}{}     & \multicolumn{1}{c}{}    & \multicolumn{1}{c}{}    & \multicolumn{1}{c}{}      &  & \multicolumn{1}{c}{}     & \multicolumn{1}{c}{}    & \multicolumn{1}{c}{}    & \multicolumn{1}{c}{}      \\
		Intercept-only & -0.049                   & 0.037                   & 0.98                    & 0.73                      &  & 0.070                    & 0.056                   & 0.98                    & 0.75                      \\
		Linear         & 0.017                    & 0.039                   & 0.99                    & 0.93                      &  & -0.006                   & 0.061                   & 0.97                    & 0.94                      \\
		Log-linear     & 0.005                    & 0.038                   & 0.99                    & 0.94                      &  & -0.004                   & 0.060                    & 0.97                    & 0.94                      \\
		Spline         & 0.001                    & 0.042                   & 1.00                    & 0.95                      &  & -0.001                   & 0.061                   & 0.97                    & 0.94                      \\
		Disjoint       & 0.000                    & 0.043                   & 1.00                    & 0.95                      &  & -0.001                   & 0.061                   & 0.97                    & 0.94                      \\
		$n=1000$       & \multicolumn{1}{c}{}     & \multicolumn{1}{c}{}    & \multicolumn{1}{c}{}    & \multicolumn{1}{c}{}      &  & \multicolumn{1}{c}{}     & \multicolumn{1}{c}{}    & \multicolumn{1}{c}{}    & \multicolumn{1}{c}{}      \\
		Intercept-only & -0.049                   & 0.026                   & 0.99                    & 0.53                      &  & 0.070                    & 0.040                    & 0.99                    & 0.57                      \\
		Linear         & 0.017                    & 0.028                   & 0.99                    & 0.90                      &  & -0.005                   & 0.042                   & 0.99                    & 0.95                      \\
		Log-linear     & 0.005                    & 0.027                   & 0.99                    & 0.94                      &  & -0.004                   & 0.042                   & 0.99                    & 0.94                      \\
		Spline         & 0.001                    & 0.030                   & 0.99                    & 0.95                      &  & 0.000                    & 0.042                   & 0.99                    & 0.95                      \\
		Disjoint       & 0.000                    & 0.031                   & 0.98                    & 0.95                      &  & 0.000                    & 0.042                   & 0.99                    & 0.95                      \\
		$n=2000$       & \multicolumn{1}{c}{}     & \multicolumn{1}{c}{}    & \multicolumn{1}{c}{}    & \multicolumn{1}{c}{}      &  & \multicolumn{1}{c}{}     & \multicolumn{1}{c}{}    & \multicolumn{1}{c}{}    & \multicolumn{1}{c}{}      \\
		Intercept-only & -0.049                   & 0.018                   & 0.99                    & 0.23                      &  & 0.070                    & 0.028                   & 0.99                    & 0.29                      \\
		Linear         & 0.017                    & 0.020                   & 0.99                    & 0.86                      &  & -0.005                   & 0.030                    & 0.99                    & 0.94                      \\
		Log-linear     & 0.005                    & 0.019                   & 0.99                    & 0.94                      &  & -0.004                   & 0.030                    & 0.99                    & 0.95                      \\
		Spline         & 0.001                    & 0.021                   & 0.99                    & 0.95                      &  & 0.000                    & 0.030                    & 0.99                    & 0.95                      \\
		Disjoint       & 0.000                    & 0.022                   & 1.00                    & 0.95                      &  & 0.000                    & 0.030                    & 0.99                    & 0.95                      \\ \hline
	\end{tabular}
	\floatfoot{ESE: empirical standard error, SER: standard error ratio, Cover: 95\% confidence interval coverage. \\
	Simulation results are for estimation of the causal risk difference with confounding due to a baseline covariate. Here, several specifications of the functional form for time were compared. In all cases, the variance for the risk difference was estimated using the empirical sandwich variance estimator. 	\\
	Bias was defined as the estimated risk difference minus the true causal risk difference. ESE was estimated by taking the standard deviation of the point estimates. SER was defined as the average standard error divided by the ESE. Confidence interval coverage was estimated by the proportion of confidence intervals that contained the true causal risk difference.\\
	Simulation metrics were computed across 5000 iterations. For $n=250$, 17 of the pooled logistic regression models with splines failed to converge in 50000 iterations of the root-finding procedure. These failures were excluded from the computation of the performance metrics. 
	}
	\label{TableSimResults}
\end{table}

\section{Conclusions}

Here, we described a novel implementation of pooled logistic regression. The proposed implementation allows for consistent variance estimation when using a pooled logistic model to estimate nuisance parameters without relying on the bootstrap. Further, the implementation described does not require the creation or storage of a long data set. \edit{Both of these aspects work to reduce the overall run-time of the g-computation estimator}, as highlighted in the applied examples. \edit{The proposed approach } allows researchers to fit pooled logistic models that otherwise might not be possible due to computational constraints, as illustrated by the second example. \edit{The exact improvements with the proposed approach will depend on the available hardware (e.g., available RAM, available cores for parallelization) and other characteristics (e.g., sample size, number of bootstrap iterations). } As of v4.1, \texttt{delicatessen} provides a built-in functionality for pooled logistic regression \edit{via the built-in \texttt{ee\_plogit} function} \cite{zivichDelicatessenMEstimationPython2022}. \edit{The approach described here can also be implemented in R using either the \texttt{geex} or \texttt{sandwich} packages \cite{saulCalculusMEstimationGeex2020}.}

While focus was on pooled logistic regression models, other pooled regression models could also be implemented \cite{craigReviewSurvivalStacking2025}. For example, complementary log-log models have been suggested as an alternative model to better approximate the hazard ratio of a Cox model \cite[pg.~240]{allisonSurvivalAnalysisUsing2010}. Alternatively, \edit{if one is interested in the cause-specific cumulative incidence function then a pooled multinomial logistic model } could be \edit{used instead in settings with } competing events, where another event precludes the event of interest from occurring \cite[pg.~251]{allisonSurvivalAnalysisUsing2010}\cite{neurothPooledMultinomialLogistic2026}. To implement these alternative pooled regression models, one can simply replace the score function of a logistic model with the score function of the desired model. Performance of estimating equations for these extensions is left for future work.

There are some limitations to the g-computation estimators that should be recognized. The pooled logistic regression model involved discretizing time. This discretization ensured the dimension of the estimating functions were finite as $n$ grows. There is also statistical theory for allowing $K$ to grow with $n$ \cite{tangLikelihoodBasedInference2012}, or for estimating functions that grow with $n$ (i.e., infinite dimension parameters) \cite{kosorokZEstimators2008}. This theory can be used to justify other estimators of $\mu_a$, such as a Cox proportional hazards model with the Breslow estimator \cite{edwardsSemiparametricGcomputationSurvival2024}. However, the conditions for these models can be more difficult to verify and some models, like the Cox model, necessitate modifications of the empirical sandwich variance estimator \cite{linRobustInferenceCox1989}. There are also alternative methods for the continuous time setting, including marginal structural Cox models, marginal structural accelerated failure time models, and structural nested failure time models. \edit{When relatively few persons are measured in the data, the empirical sandwich variance estimator can be biased. A variety of different finite-sample corrections have been proposed for the empirical sandwich variance estimator \cite{saulCalculusMEstimationGeex2020, nomaRobustVarianceEstimators2026}, which could be adapted to the setting considered here in principle. } Another limitation is that the described implementation does not allow for time-varying treatments (i.e., when $A$ is a time-varying variable). In these settings, there are concerns over time-varying confounding, where a variable is affected by prior treatment but affects future treatments. Standard regression methods fail to estimate causally interpretable parameters in these settings \cite{robinsEstimationCausalEffect1999}. Instead, one can use the iterated conditional expectation g-formula \cite{wenParametricGformulaImplementations2021}, which has been previously expressed using estimating equations \cite{zivichEmpiricalSandwichVariance2024}, but this approach uses a series of nested regression models. Another popular approach for time-varying actions is a censor-weight approach \cite{robinsCorrectingNoncomplianceDependent2000a}, which can rely on a pooled logistic model. Extending the proposed estimating equations to allow the covariate design matrix to vary over time would then allow for the empirical sandwich variance estimator to be used for inference in this setting. This extension is left for future work.

\section*{Acknowledgments}

Conflicts of Interest: None to declare. 
~\\~\\
Funding: This work was supported in part by K01AI177102 (PNZ), R01AI157758 (PNZ, SRC, JKE, BES), K01AI182506 (BES), R37AI054165 (JBD), P30AI050410 (PNZ, SRC, BES). The content is solely the responsibility of the authors and does not necessarily represent the official views of the National Institutes of Health.
~\\~\\
Data Availability: Data and code used for the illustrative examples and the simulation experiment are publicly available on GitHub at \url{https://github.com/pzivich/publications-code} under \texttt{PooledLogitEE/} or via Zenodo 10.5281/zenodo.15333626.

\bibliography{references}{}
\bibliographystyle{ieeetr}

\newpage

\section*{Appendix A}

Here, the closed-form analytic Jacobian, $\nabla_{\theta} \psi(O_i; \theta)$ for Equation 6. The Jacobian of the stacked estimating functions for $\psi_r$ can be expressed as
\begin{equation*}
	\begin{bmatrix}
		\nabla_{\beta} \psi(O_i; \beta) & \frac{\partial}{\partial \gamma^a_m} \psi(O_i; \beta) \\
		\nabla_{\beta} \psi_r(O_i; \beta, \gamma^a_m) & \frac{\partial}{\partial \gamma^a_m} \psi_r(O_i; \beta, \gamma^a_m) \\		
	\end{bmatrix}
\end{equation*} 
Starting with the upper left, first recall that 
$$\frac{\partial}{\partial z} \text{expit}(z) = \frac{\partial}{\partial z} \left(\frac{1}{1 + \exp(-z)}\right) = \text{expit}(z) \times \left\{ 1 - \text{expit}(z) \right\}.$$
Therefore, for general element $x_i \in \mathbb{X}_i$ with the corresponding parameter $\beta_x$, it follows that
\begin{equation*}
	\frac{\partial}{\partial x} \psi(O_i; \beta) = - \sum_{k \in \mathcal{K}} \left\{ I(T_i^* > s_{k-1}) \hat{Y}_{i,k} \left( 1 - \hat{Y}_{i,k} \right) x_i^2
	\right\}
\end{equation*}
from the sum and product rules and the equality given for $\text{expit}$. This process can be repeated for all design matrix components in $\mathbb{X}_i$, as well as $\mathbb{S}_{i_k}$, and their corresponding parameters to obtain $\nabla_{\beta} \psi(O_i; \beta)$. Note that it is straightforward to show that $\frac{\partial}{\partial \gamma^a_m} \psi(O_i; \beta) = 0$ since $\psi(O_i; \beta)$ does not involve the parameter $\gamma^a_m$. It is similarly straightforward to show that $\frac{\partial}{\partial \gamma^a_m} \psi_r(O_i; \theta) = -1$. Again, letting $x_i$ denote a general element in the design matrix and $\beta_x$ denote the corresponding parameter, now consider $\frac{\partial}{\partial \beta_x} \psi_r(O_i; \beta, \gamma^a_m)$. Notice that $\frac{\partial}{\partial \beta_x} \hat{Y}_{i,k} = \hat{Y}_{i,k} \left( 1 - \hat{Y}_{i,k} \right) x_i$ and the derivative of the cumulative product is $ \frac{\partial}{\partial z} \prod_{j=1}^k f_j(z) = \sum_{j=1}^{k} \left\{ f_j'(z) \prod_{l \ne j} f_l(z) \right\} $ via the chain rule. Therefore, 
\begin{equation*}
	\frac{\partial}{\partial \beta_x} \psi_r(O_i; \beta, \gamma^a_m) 
	= \sum_{j=1}^{k} \left\{ \hat{Y}_{i,j} \left( 1 - \hat{Y}_{i,j} \right) x_i \prod_{l \ne j} (1 - \hat{Y}_{i,l}) \right\}
	= \sum_{j=1}^{k} \left\{ \hat{Y}_{i,j} x_i \prod_{j}^k (1 - \hat{Y}_{i,j}) \right\}
	= \gamma^a_m \sum_{j=1}^{k} \left\{ \hat{Y}_{i,j} x_i \right\}
\end{equation*}
following from the reviewed equalities, moving the multiplication for $j$ into the cumulative product, and the definition of $\gamma^a_m$, respectively. As before, this process applies to all design matrix components in $\mathbb{X}_i$ and $\mathbb{S}_{i_k}$. From these elements, the Jacobian can be constructed.

\newpage

\section*{Appendix B}

The following is an illustrative example of the computations for the described vectorization of the estimating equations. For $5$ units, let the observed times be $[1, 2, 2, 4, 4, 5]$ and the event indicators be $[1, 1, 0, 1, 0, 0]$. Recall that person-period observations are the corresponding row in a long data set that corresponds to a unit's contribution for each discrete time interval. Additionally, consider a single baseline covariate with values of $[-1, 1, -1, 0, 2, -2]$. Then
\begin{equation*}
	\mathbb{X}_{p \times n} = 
	\begin{bmatrix}
		-1 \\
		1 \\
		-1 \\
		0 \\
		2 \\ 
		-2
	\end{bmatrix} 
\end{equation*}
For $\beta_X = 0.2$, $(\mathbb{X}_{p \times n} \cdot \beta_X)^T = (-0.2, 0.2, -0.2, 0, 0.4, -0.4)$ and
\begin{equation*}
	\mathcal{X} = 
	\begin{bmatrix}
		-0.2 & 0.2 & -0.2 & 0 & 0.4 & -0.4 \\
		-0.2 & 0.2 & -0.2 & 0 & 0.4 & -0.4 \\
		-0.2 & 0.2 & -0.2 & 0 & 0.4 & -0.4 \\
		-0.2 & 0.2 & -0.2 & 0 & 0.4 & -0.4 \\
		-0.2 & 0.2 & -0.2 & 0 & 0.4 & -0.4 \\
	\end{bmatrix} .
\end{equation*}
Below is Python code to setup the following example
\begin{python}
	import numpy as np
	from delicatessen.utilities import inverse_logit
	
	t = np.asarray([1, 2, 2, 4, 4, 5])
	d = np.asarray([1, 1, 0, 1, 0, 0])
	X = np.asarray([[-1, 1, -1, 0, 2, -2], ]).T
	beta_x = [0.2, ]
\end{python}

\subsection*{Smooth Functional Form for Time}
Consider the case where time is modeled as a linear function with $\beta_S = (-1, 0.1)$. Therefore,
\begin{equation*}
	\mathcal{S} = 
	\begin{bmatrix}
		1 & 1 \\
		1 & 2 \\
		1 & 3 \\
		1 & 4 \\
		1 & 5 \\
	\end{bmatrix} 
\end{equation*}
\begin{equation*}
	\mathbf{\Delta} = 
	\begin{bmatrix}
		1 & 1 & 0 & 1 & 0 & 0 \\
		1 & 1 & 0 & 1 & 0 & 0 \\
		1 & 1 & 0 & 1 & 0 & 0 \\
		1 & 1 & 0 & 1 & 0 & 0 \\
		1 & 1 & 0 & 1 & 0 & 0 \\
	\end{bmatrix} 
\end{equation*}
and
\begin{equation*}
	\mathcal{S} \cdot \beta_S^T = 
	\begin{bmatrix}
		1 & 1 \\
		1 & 2 \\
		1 & 3 \\
		1 & 4 \\
		1 & 5 \\
	\end{bmatrix} \cdot 
	\begin{bmatrix}
		-1 \\ 
		0.1 \\
	\end{bmatrix} 
	= 	
	\begin{bmatrix}
		-0.9 \\
		-0.8 \\
		-0.7 \\
		-0.6 \\
		-0.5 \\		
	\end{bmatrix}
\end{equation*}
Based on the times and event indicators, the risk set and final time under observation matrices are defined as 
\begin{equation*}
	\mathcal{R} = 
	\begin{bmatrix}
		1 & 1 & 1 & 1 & 1 & 1 \\
		0 & 1 & 1 & 1 & 1 & 1 \\
		0 & 0 & 0 & 1 & 1 & 1 \\
		0 & 0 & 0 & 1 & 1 & 1 \\
		0 & 0 & 0 & 0 & 0 & 1 \\
	\end{bmatrix} 
	\qquad \qquad \qquad \qquad \qquad
	\mathcal{R}^* = 
	\begin{bmatrix}
		1 & 0 & 0 & 0 & 0 & 0 \\
		0 & 1 & 1 & 0 & 0 & 0 \\
		0 & 0 & 0 & 0 & 0 & 0 \\
		0 & 0 & 0 & 1 & 1 & 0 \\
		0 & 0 & 0 & 0 & 0 & 1 \\
	\end{bmatrix} 
\end{equation*}	
Therefore,
\begin{equation*}
	\mathcal{Y} = \mathbf{\Delta} \odot \mathcal{R}^* =
	\begin{bmatrix}
		1 & 0 & 0 & 0 & 0 & 0 \\
		0 & 1 & 0 & 0 & 0 & 0 \\
		0 & 0 & 0 & 0 & 0 & 0 \\
		0 & 0 & 0 & 1 & 0 & 0 \\
		0 & 0 & 0 & 0 & 0 & 0 \\
	\end{bmatrix}	
\end{equation*}
and
\begin{equation*}
	\hat{\mathcal{Y}} = \text{expit} \left(
	\begin{bmatrix}
		-0.2 & 0.2 & -0.2 & 0 & 0.4 & -0.4 \\
		-0.2 & 0.2 & -0.2 & 0 & 0.4 & -0.4 \\
		-0.2 & 0.2 & -0.2 & 0 & 0.4 & -0.4 \\
		-0.2 & 0.2 & -0.2 & 0 & 0.4 & -0.4 \\
		-0.2 & 0.2 & -0.2 & 0 & 0.4 & -0.4 \\
	\end{bmatrix} \oplus	
	\begin{bmatrix}
		-0.9 \\
		-0.8 \\
		-0.7 \\
		-0.6 \\
		-0.5 \\		
	\end{bmatrix} 
	\right) =
	\begin{bmatrix}
		0.250 & 0.332 & 0.250 & 0.289 & 0.378 & 0.214 \\
		0.269 & 0.354 & 0.269 & 0.310 & 0.401 & 0.231 \\
		0.289 & 0.378 & 0.289 & 0.332 & 0.426 & 0.250 \\
		0.310 & 0.401 & 0.310 & 0.354 & 0.450 & 0.269 \\
		0.332 & 0.426 & 0.332 & 0.378 & 0.475 & 0.289 \\
	\end{bmatrix}
\end{equation*}	
which gives 
\begin{equation*}
	\mathcal{P} = (\mathcal{Y} - \hat{\mathcal{Y}}) \odot \mathcal{R} =
	\begin{bmatrix}
		0.750 & -0.332 & -0.250 & -0.289 & -0.378 & -0.214 \\
		0     &  0.646 & -0.269 & -0.310 & -0.401 & -0.231 \\
		0     &  0     &  0     & -0.332 & -0.426 & -0.250 \\
		0     &  0     &  0     &  0.646 & -0.450 & -0.269 \\
		0     &  0     &  0     &  0     &  0     & -0.289 \\
	\end{bmatrix}
\end{equation*}
So, the contributions to the estimating equations for the covariates $X$ can be computed via
\begin{equation*}
	\left(\mathbf{1}_{1 \times K} \cdot \mathcal{P}\right)^T = 
	\left(
	\begin{bmatrix}
		1 & 1 & 1 & 1 & 1
	\end{bmatrix}
	\begin{bmatrix}
		0.750 & -0.332 & -0.250 & -0.289 & -0.378 & -0.214 \\
		0     &  0.646 & -0.269 & -0.310 & -0.401 & -0.231 \\
		0     &  0     &  0     & -0.332 & -0.426 & -0.250 \\
		0     &  0     &  0     &  0.646 & -0.450 & -0.269 \\
		0     &  0     &  0     &  0     &  0     & -0.289 \\
	\end{bmatrix}
	\right)^T =
	\begin{bmatrix}
		0.750 \\
		0.314 \\
		-0.519 \\
		-0.285 \\
		-1.655 \\
		-1.253 \\
	\end{bmatrix}
\end{equation*}
so
\begin{equation*}
	(\mathbf{1}_{1 \times K} \cdot \mathcal{P})^T \odot \mathbb{X} =
	\begin{bmatrix}
		0.750 \\
		0.314 \\
		-0.519 \\
		-0.285 \\
		-1.655 \\
		-1.253 \\
	\end{bmatrix}
	\odot 	
	\begin{bmatrix}
		-1 \\
		1 \\
		-1 \\
		0 \\
		2 \\ 
		-2
	\end{bmatrix} 
	= 
	\begin{bmatrix}
		-0.750 \\
		0.314 \\
		0.519 \\
		0     \\
		-3.309 \\ 
		2.507
	\end{bmatrix} 
\end{equation*}
The score function for time is then
\begin{equation*}
	\mathcal{S}^T \cdot \mathcal{P} = 
	\begin{bmatrix}
		1 & 1 & 1 & 1 & 1 \\
		1 & 2 & 3 & 4 & 5 \\
	\end{bmatrix} \cdot 
	\begin{bmatrix}
		0.750 & -0.332 & -0.250 & -0.289 & -0.378 & -0.214 \\
		0     &  0.646 & -0.269 & -0.310 & -0.401 & -0.231 \\
		0     &  0     &  0     & -0.332 & -0.426 & -0.250 \\
		0     &  0     &  0     &  0.646 & -0.450 & -0.269 \\
		0     &  0     &  0     &  0     &  0     & -0.289 \\
	\end{bmatrix}
\end{equation*}
\begin{equation*}
	=
	\begin{bmatrix}
		0.750 & 0.314 & -0.519 & -0.285 & -1.655 & -1.253 \\
		0.750 & 0.960 & -0.788 & -0.678 & -4.258 & -3.947 \\		
	\end{bmatrix}
\end{equation*}
This gives the final stacked estimating functions of
\begin{equation*}
	\begin{bmatrix}
		-0.750 & 0.314 &  0.519 &  0     & -3.309 &  2.507 \\
		0.750 & 0.314 & -0.519 & -0.285 & -1.655 & -1.253 \\
		0.750 & 0.960 & -0.788 & -0.678 & -4.258 & -3.947 \\		
	\end{bmatrix}
\end{equation*}
The following Python code implements the described steps
\begin{python}
	beta_s = [-1, 0.1]
	time_steps = np.asarray(range(1, 6))
	n_time_steps = len(time_steps)
	
	# \mathcal{X}
	log_odds_x = np.dot(X, beta_x)
	log_odds_x_matrix = np.tile(log_odds_x, (n_time_steps, 1))  #
	
	# \mathcal{S}
	intercept = np.ones(time_steps.shape)[:, None]
	S = np.concatenate([intercept, time_steps[:, None]], axis=1)
	log_odds_t = np.dot(S, beta_s)
	
	# \mathcal{R}
	risk_set = (t >= time_steps[:, None]).astype(int)
	
	# \mathcal{R}^*
	last_time = (t == time_steps[:, None]).astype(int)
	
	# \mathcal{Y}
	y_obs = d * last_time
	
	# \hat{\mathcal{Y}}
	y_pred = inverse_logit(log_odds_x_matrix + log_odds_t[:, None])
	
	# \mathcal{P}
	residual_matrix = (y_obs - y_pred) * risk_set
	
	# Contributions by X
	n_ones = np.ones(shape=(1, n_time_steps))
	y_resid = np.dot(n_ones, residual_matrix).T
	x_score = y_resid * X
	
	# Contributions by S
	t_score = np.dot(S.T, residual_matrix)
	
	# Estimating functions
	est_func = np.vstack([x_score.T, t_score])
\end{python}
The estimating function can also be evaluated using the \texttt{ee\_plogit} function from \texttt{delicatessen} in Python via
\begin{python}
	# Creating S design matrix
	time_steps = np.asarray(range(1, 6))
	n_time_steps = len(time_steps)
	intercept = np.ones(time_steps.shape)[:, None]
	s_matrix = np.concatenate([intercept, time_steps[:, None], ], axis=1)
	
	# Evaluating the estimating functions at beta
	ee_plogit(theta=beta_x + beta_s, X=X, t=t, delta=d, S=s_matrix)
\end{python}

\subsection*{Disjoint Indicators for Time}

Now consider the modified procedure for disjoint indicators. The unique event times are $(1, 2, 4)$. Let $\beta_S = (-1, 0.1, -0.1)$. So,
\begin{equation*}
	\mathcal{S} = 
	\begin{bmatrix}
		1 & 0 & 0 \\
		1 & 1 & 0 \\
		1 & 0 & 1 \\
	\end{bmatrix} 
\end{equation*}
\begin{equation*}
	\mathbf{\Delta} = 
	\begin{bmatrix}
		1 & 1 & 0 & 1 & 0 & 0 \\
		1 & 1 & 0 & 1 & 0 & 0 \\
		1 & 1 & 0 & 1 & 0 & 0 \\
	\end{bmatrix} 
\end{equation*}
and 
\begin{equation*}
	\mathcal{S} \cdot \beta_S^T = 
	\begin{bmatrix}
		1 & 0 & 0 \\
		1 & 1 & 0 \\
		1 & 0 & 1 \\
	\end{bmatrix} 
	\begin{bmatrix}
		-1  \\
		0.1 \\
		-0.1 
	\end{bmatrix} 
	=
	\begin{bmatrix}
		-1 \\
		-0.9 \\
		-1.1
	\end{bmatrix}
\end{equation*}
The matrices for the risk set and final time under observation are then
\begin{equation*}
	\mathcal{R} = 
	\begin{bmatrix}
		1 & 1 & 1 & 1 & 1 & 1 \\
		0 & 1 & 1 & 1 & 1 & 1 \\
		0 & 0 & 0 & 1 & 1 & 1 \\
	\end{bmatrix} 
	\qquad \qquad \qquad \qquad \qquad
	\mathcal{R}^* = 
	\begin{bmatrix}
		1 & 0 & 0 & 0 & 0 & 0 \\
		0 & 1 & 1 & 0 & 0 & 0 \\
		0 & 0 & 0 & 1 & 1 & 0 \\
	\end{bmatrix} 
\end{equation*}	
Thus,
\begin{equation*}
	\mathcal{Y} = \mathbf{\Delta} \odot \mathcal{R}^* = 
	\begin{bmatrix}
		1 & 0 & 0 & 0 & 0 & 0 \\
		0 & 1 & 0 & 0 & 0 & 0 \\
		0 & 0 & 0 & 1 & 0 & 0 \\
	\end{bmatrix} 
\end{equation*}
and
\begin{equation*}
	\hat{\mathcal{Y}} =  
	\begin{bmatrix}
		0.231 & 0.310 & 0.231 & 0.269 & 0.354 & 0.198 \\
		0.250 & 0.332 & 0.250 & 0.289 & 0.378 & 0.214 \\
		0.214 & 0.289 & 0.214 & 0.250 & 0.332 & 0.182 \\
	\end{bmatrix} 
\end{equation*}
Then
\begin{equation*}
	\mathcal{P} = 
	\begin{bmatrix}
		0.769 & -0.310 & -0.231 & -0.269 & -0.354 & -0.198 \\
		0     &  0.668 & -0.250 & -0.289 & -0.378 & -0.214 \\
		0     &  0.    &  0     &  0.750 & -0.332 & -0.182 \\
	\end{bmatrix} 	
\end{equation*}
Therefore, the individual contributions to the estimating equations for $X$ are
\begin{equation*}
	(\mathbf{1}_{1 \times K^*} \cdot \mathcal{P})^T \odot \mathbb{X} =
	\begin{bmatrix}
		0.769 \\
		0.358 \\
		-0.481 \\
		0.192 \\
		-1.064 \\
		-0.594 \\
	\end{bmatrix}
	\odot 	
	\begin{bmatrix}
		-1 \\
		1 \\
		-1 \\
		0 \\
		2 \\ 
		-2
	\end{bmatrix} 
	= 
	\begin{bmatrix}
		-0.769 \\
		0.358 \\
		0.418 \\
		0     \\
		-2.127 \\ 
		1.189
	\end{bmatrix} 
\end{equation*}
As stated in the manuscript, the individual contributions to the estimating equations for time with disjoint indicators is simply $\mathcal{P}$. So, the final stacked estimating functions are
\begin{equation*}
	\begin{bmatrix}
		-0.769 &  0.358 &  0.481 &  0     & -2.127 &  1.189 \\
		0.769 & -0.310 & -0.231 & -0.269 & -0.354 & -0.198 \\
		0     &  0.668 & -0.250 & -0.289 & -0.378 & -0.214 \\
		0     &  0.    &  0     &  0.750 & -0.332 & -0.182 \\
	\end{bmatrix}
\end{equation*}
The following Python code implements the described steps
\begin{python}
	beta_s = [-1, 0.1, -0.1]
	event_times = t[d == 1]
	unique_event_times = np.unique(event_times)
	tp = unique_event_times.shape[0]
	
	# \mathcal{X}
	log_odds_x = np.dot(X, beta_x)
	log_odds_x_matrix = np.tile(log_odds_x, (tp, 1))
	
	# \mathcal{S}
	time_design_matrix = np.identity(n=len(unique_event_times))
	time_design_matrix[:, 0] = 1
	log_odds_t = np.dot(time_design_matrix, beta_s)
	
	# \mathcal{R}
	risk_set = (t >= unique_event_times[:, None]).astype(int)
	
	# \mathcal{R}^*
	last_time = (t == unique_event_times[:, None]).astype(int)
	
	# \mathcal{Y}
	y_obs = d * last_time
	
	# \hat{\mathcal{Y}}
	y_pred = inverse_logit(log_odds_x_matrix + log_odds_t[:, None])
	
	# \mathcal{P}
	residual_matrix = (y_obs - y_pred) * risk_set
	
	# Contributions by X
	n_ones = np.ones(shape=(1, tp))
	y_resid = np.dot(n_ones, residual_matrix)[0]
	x_score = y_resid[:, None] * X
	
	# Contributions by S
	t_score = residual_matrix
	
	# Estimating functions
	est_func = np.vstack([x_score.T, t_score])
\end{python}
Again, the estimating function can be evaluated using the \texttt{ee\_plogit} interface available in \texttt{delicatessen} via
\begin{python}	
	# Evaluating the estimating functions at beta
	ee_plogit(theta=beta_x + beta_s, X=X, t=t, delta=d)
\end{python}

\newpage 

\section*{Appendix C}

\begin{figure}[h]
	\centering
	\caption{Bias for the causal risk difference at selected time points by sample size and functional form for time}
	\includegraphics[width=0.85\linewidth]{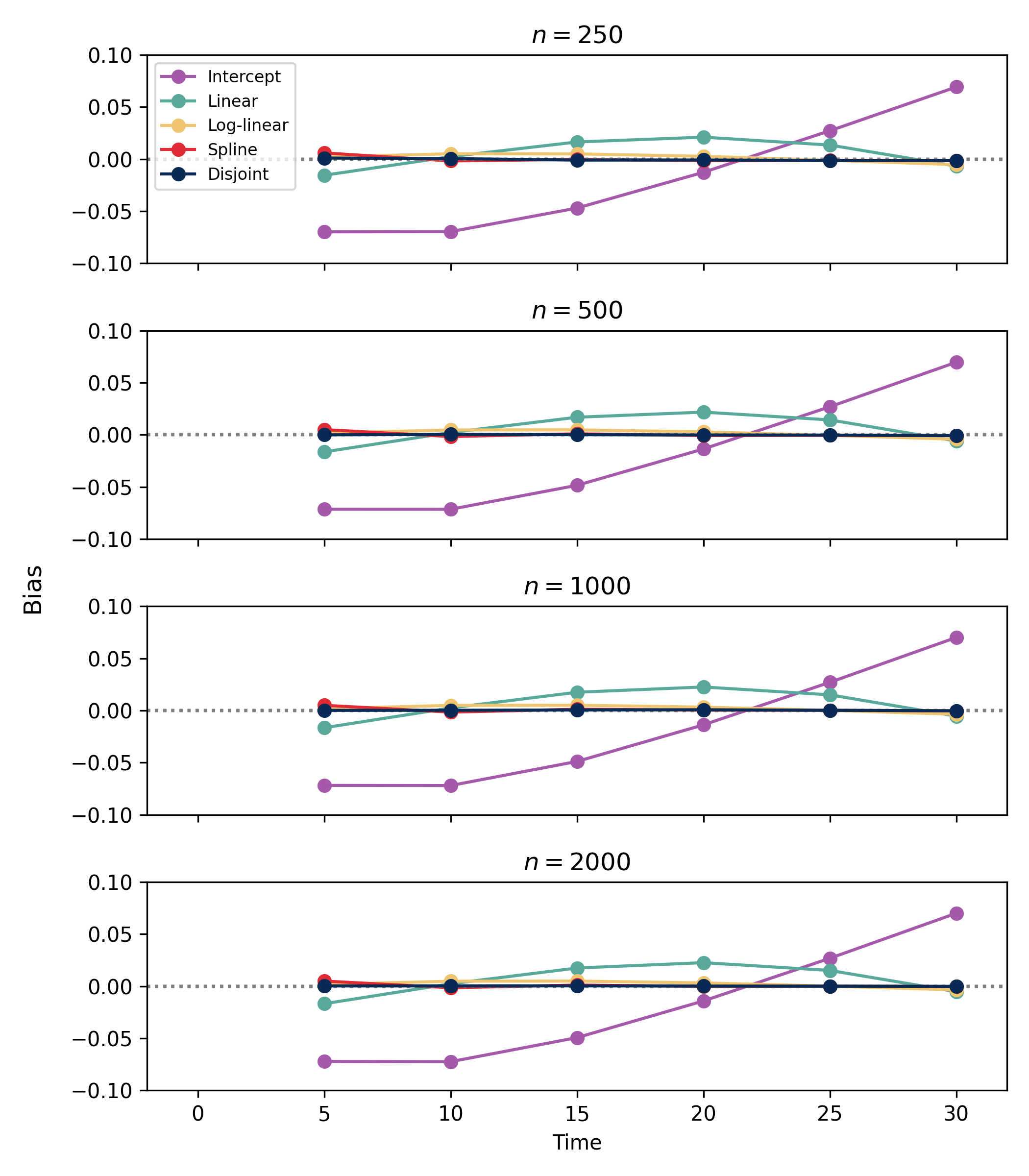} 
	\label{FigA1}
\end{figure}

\begin{figure}
	\centering
	\caption{95\% confidence interval coverage for the causal risk difference at selected time points by sample size and functional form for time}
	\includegraphics[width=0.85\linewidth]{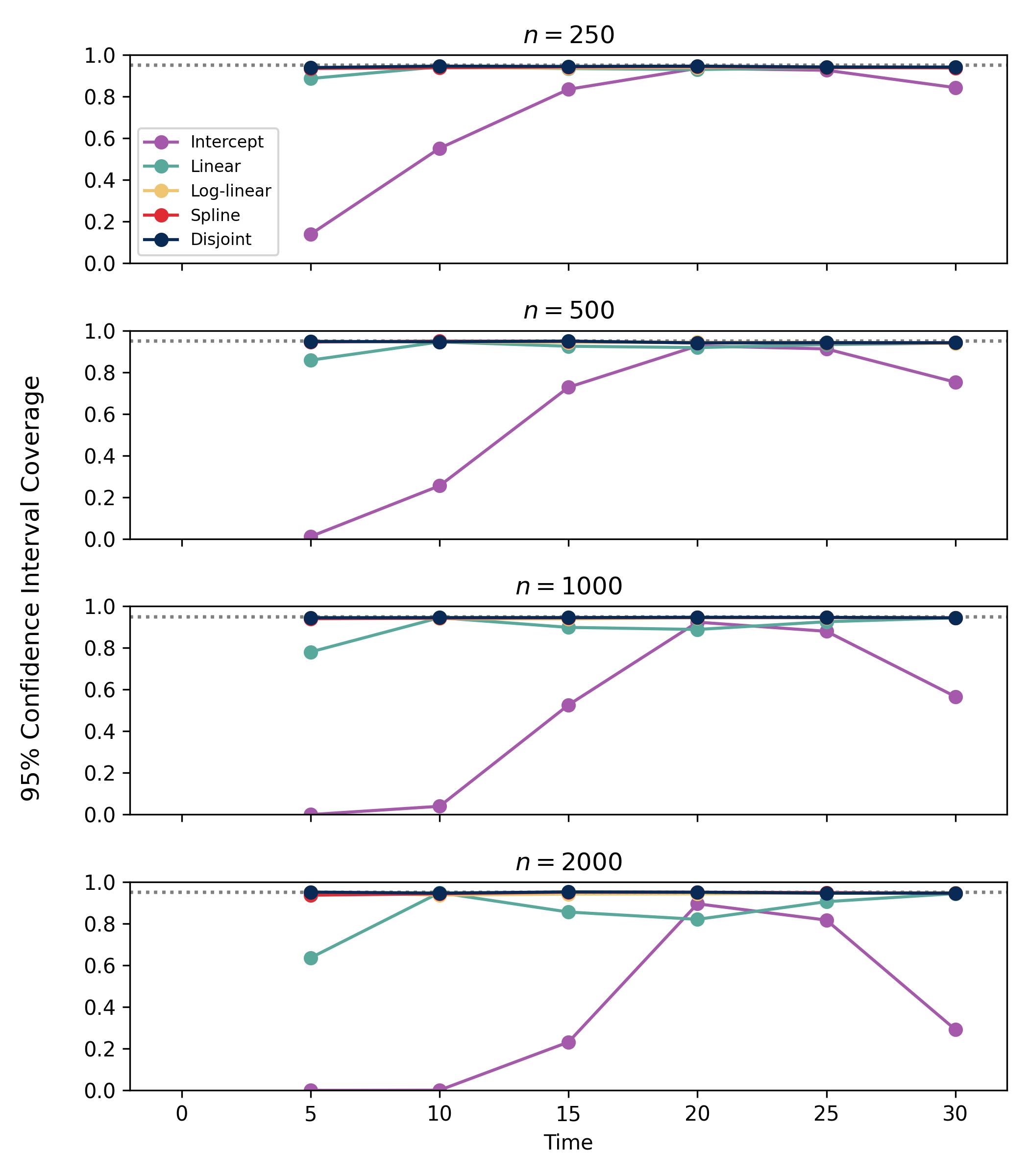} 
	\label{FigA2}
\end{figure}

\end{document}